\documentclass[12pt,preprint]{aastex}
\usepackage{color}
\usepackage{graphicx}

\shorttitle{DeFN-R model for Solar Flare Prediction}
\shortauthors{N. Nishizuka et al.}

\begin{document}

\title{RELIABLE PROBABILITY FORECAST OF SOLAR FLARES: DEEP FLARE NET-RELIABLE (DeFN-R)}


\author{Naoto Nishizuka\altaffilmark{1}, Y\^uki Kubo\altaffilmark{1}, Komei Sugiura\altaffilmark{2,3}, Mitsue Den\altaffilmark{1} and Mamoru Ishii\altaffilmark{1}}
\altaffiltext{1}{Applied Electromagnetic Research Institute, National Institute of Information and Communications Technology, 4-2-1, Nukui-Kitamachi, Koganei, Tokyo 184-8795, 
Japan; nishizuka.naoto@nict.go.jp}
\altaffiltext{2}{Department of Information and Computer Science, Keio University, 3-14-1 Hiyoshi, Kohoku, Yokohama, Kanagawa 223-8522, Japan}
\altaffiltext{3}{Advanced Speech Translation Research and Development Promotion Center, National Institute of Information and Communications Technology}

\begin{abstract}
We developed a reliable probabilistic solar flare forecasting model using a deep neural network, named Deep Flare Net-Reliable (DeFN-R). The model can predict the maximum classes 
of flares that occur in the following 24 h after observing images, along with the event occurrence probability. We detected active regions from 3$\times$10$^5$ solar images taken 
during 2010--2015 by Solar Dynamic Observatory and extracted 79 features for each region, which we annotated with flare occurrence labels of X-, M-, and C-classes. The extracted 
features are the same as used by Nishizuka et al. (2018); for example, line-of-sight/vector magnetograms in the photosphere, brightening in the corona, and the X-ray emissivity 1 and 
2 h before an image. We adopted a chronological split of the database into two for training and testing in an operational setting: the dataset in 2010--2014 for training and the one in 
2015 for testing. DeFN-R is composed of multilayer perceptrons formed by batch normalizations and skip connections. By tuning optimization methods, DeFN-R was trained to optimize 
the Brier skill score (BSS). As a result, we achieved BSS = 0.41 for $\geq$C-class flare predictions and 0.30 for $\geq$M-class flare predictions by improving the reliability diagram 
while keeping the relative operating characteristic curve almost the same. Note that DeFN is optimized for deterministic prediction, which is determined with a normalized threshold of 
50\%. On the other hand, DeFN-R is optimized for a probability forecast based on the observation event rate, whose probability threshold can be selected according to users' purposes.

\end{abstract}


%
\section{Introduction}

Solar flare prediction is a long-standing problem in solar physics and astrophysics. The fundamental physical process of solar flares has been unveiled by recent theoretical and 
observational studies, but neither a standard approach to predict flares nor understanding of the mechanism of flare occurrence has been established. There are four approaches 
to flare prediction: (i) empirical human forecasting, (ii) statistical method, (iii) machine learning method, and (iv) numerical simulation by solving physical equations. Each model has 
been developed to predict categories or occurrence probabilities of flares in the following 24 h. Since the number of large-class flares is much smaller than that of small-class 
flares or non-flare events, flare prediction is an imbalanced problem.

Probability forecasts are widely used in human forecasting \citep{cro12, dev14, murr17}, as well as in statistical methods \citep{whe00, moo01, gal02, whe05, lee12, blo12, blo16, 
lee16, mcc16, mcc18, lek18, lim19a, lim19b}. It is more convenient to show the occurrence probability for each class of flares rather than the deterministic one when the observation 
is near the border of two categories. Ideally, the forecast probability is equal to the observation event rate. The prediction result is often evaluated by the relative (receiver) operating 
characteristic (ROC) curve \citep{swe73, mas82, sta89, har92}, the reliability diagram, and the Brier skill score (BSS). BSS uses the climatological event rate as a reference forecast 
and is complemented by a reliability diagram that compares the forecast probabilities and the observation frequency. Barnes (2016) compared eleven probability forecasting models of 
solar flares and concluded that several of them lack reliability \citep[see also][]{lek19, par20}.

On the other hand, it is currently popular to apply machine learning algorithms to predict categories of flares occurring in the following 24 h. In previous works, classical algorithms 
such as neural networks \citep{qah07, col09, hig11, ahm13}, support vector machines \citep[SVMs;][]{qah07, yua10, alg15, bob15, bou15, mur15, rab17, sad17, ali19}, the least absolute 
shrinkage and selection operator \citep[LASSO;][]{ben18, jon18}, random forests \citep[RFs;][]{liu17, flo18, wan19, cam19}, extremely randomized trees, which is an extension of the 
RF model with higher accuracy \citep[ERTs;][]{nis17}, and an unsupervised fuzzy clustering \citep{ben18} were applied to flare prediction. They are trained to increase a skill score, 
such as the true skill statistic (TSS), also known as the Peirce skill score, or the Hanssen and Kuipers discriminant (H\&KSS). TSS has an advantage of not depending on the ratio 
of positive and negative events in the database, as well as the odds ratio skill score and the symmetric extremal dependence index. Furthermore, deep neural networks (DNNs) have 
also recently been applied and have succeeded in improving TSS \citep{hua18, nis18, par18, liu19, che19, zhe19, dom19, yi20, li20, pan20}. 

Machine-learning-based forecast algorithms including DNNs, which were trained to increase TSS, show forecast results with higher TSS values but lower BSS values and lack 
reliability \citep{bar16}. The reliability diagrams show a systematic over-forecast for larger occurrence probability. Unreliable probabilistic forecast models are often calibrated by 
the terrestrial weather forecast community \citep[e.g.,][]{gne07, pri09}, but such calibration is not popular in the space weather forecast community. Instead, each model is requested 
to directly forecast solar flares reliably. However, a useful optimization method employing machine learning algorithms to realize the probability forecast of solar flares with high 
reliability has not yet been discussed. It has been mathematically verified that for reliable prediction models, TSS is maximized at the probability threshold of the climatological 
event rate, while the converse is not obvious \citep{kub19}.

In this study, we investigated optimization methods to realize reliable probabilistic forecasts by machine learning algorithms. We improved our prediction model of solar flares using 
DNNs, named Deep Flare Net \citep[DeFN;][]{nis18} to obtain reliable probability forecasts, and we renamed it Deep Flare Net-Reliable (DeFN-R). In sections 2 and 3, we describe 
our model and datasets. The prediction results are explained in section 4 and summarized in section 5, with discussion of the optimization method suitable for producing reliable 
models.

\clearpage
%
%
\section{Overview of DeFN-R}   
\subsection{General Explanation of Model}   

We used a solar flare prediction model employing DNNs named DeFN to investigate how to realize reliable probabilistic forecasts \citep{nis18}. We performed probabilistic forecasts 
for two categories: (i) $\geq$M-class and $<$M-class/non-flare events and (ii) $\geq$C-class and $<$C-class/non-flare events. We used the database of observation data taken 
by Geostationary Operational Environmental Satellite (GOES) and Solar Dynamic Observatory \citep[SDO;][]{pes12} from June 2010 to December 2015. During this period, 90\% 
of the observed flares, such as 26 X-class, 383 M-class, and 4054 C-class flares, occurred on the solar disk. Since the number of X-class flares was too small to evaluate the 
prediction results, X-class flares were not predicted in this paper.

Occurrence probabilities of flares are predicted by the following steps. (i) First, we download observation data from the data archives of SDO and GOES, such as the light curves 
of the soft X-ray intensity, line-of-sight magnetograms, vector magnetograms, and images obtained by the 1600 \,\AA\ and 131 \,\AA\ filters. (ii) Second, active regions (ARs) are 
detected from the full-disk line-of-sight magnetograms and numbered by tracking their time evolution. (iii) For each AR, physics-based features are extracted from multiwavelength 
observation images, and the feature database is annotated with flare labels if an X-, M-, or C-class flare occurs within 24 h after obtaining an image. (iv) The flare occurrence 
probabilities in the following 24 h is predicted by supervised machine learning using a DNN for each region with a 1 h cadence. Because we investigated reliable outputs by tuning 
DeFN, we rename it DeFN-Reliable (DeFN-R) in this paper.

\clearpage
%
%
\subsection{Observation Data and AR Features}   

First, we made the database from 3$\times$10$^5$ solar images of the full-disk line-of-sight magnetograms obtained by the Helioseismic and Magnetic Imager \citep[HMI;][]{sche12, 
scho12, hoe14} on board SDO, from which we automatically detected ARs with a reduced cadence of 1 h \citep[see details of the detection rules described in][]{nis17}. We neglected 
ARs on the limb and numbered the same ARs by tracking their time evolution. To extract various kinds of features, the frame coordinates of the detected ARs were applied to other 
observation images with different wavelengths, such as vector magnetograms taken by HMI/SDO and 1600 \,\AA\ and 131 \,\AA\ filter images by the Atmospheric Imaging Assembly 
\citep[AIA;][]{lem12} on board SDO.

Next we calculated 79 features from each AR observed by multiwavelength emissions. The features are parameters based on physics that were considered to be effective for solar 
flare prediction in previous papers and in daily forecasting operations. The extracted features are the same as used by Nishizuka et al. (2018); for example, line-of-sight/vector 
magnetograms in the photosphere, brightening in the corona at 131 \,\AA\ (T$\geq$10$^7$ K), and the 131 \,\AA\ and X-ray emissivity 1 and 2 h before an image. The sunspot 
area, the magnetic flux, the number of magnetic neutral lines, and the Lorentz force were calculated from the line-of-sight/vector magnetograms. Preflare brightening in the bottom 
chromosphere was detected in the UV--continuum taken by the 1600 \,\AA\ filter, and the coronal heating in the flaring region over 10$^7$ K was taken by the 131 \,\AA\ filter, 
as well as the integrated X-ray emission in the range of 1-8 \,\AA\ observed by GOES.

Furthermore, we standardized the feature database before the input into the DNN \citep[e.g.,][]{bis06}. The database of 2010--2015 was chronologically split into two: the dataset 
in 2010--2014 for training and the one in 2015 for validation and testing. These chronological datasets for training and testing are more challenging for predicting flares than the 
randomly shuffled and divided datasets \citep[e.g.,][]{bob15, mur15, nis17, nis18}. The optimization and evaluations were repeated several times, using subsets of the training and 
testing datasets.

\clearpage
%
%
\section{Details of DeFN-R}
\subsection{Architecture}

DeFN-R is based on algorithms of deep neural networks and composed of multilayer perceptrons. The input is the database of 79 features standardized in the preprocess, i.e., 
79-dimensional vectors of features. The output is the forecast probability of each class of flares, $p(y)$. Here, $y$ = (0, 1) is the class of $\geq$M-class flare events and $y$ = 
(1, 0) is the one of $<$M-class or non-flare events. We calculate the two probabilities of $\geq$M-class flare events and $<$M-class or non-flare events.

The architecture of DeFN-R is fundamentally the same as that of DeFN. DeFN-R consists of several layers. At each layer of the neural networks, the input is converted to the 
output with a linear combination and an activation function. As the activation function, we used the rectified linear units \citep[ReLU;][]{nai10} for the first layer to the penultimate 
layer and a softmax function for the last layer,
\begin{equation}
Softmax( x_i ) = \frac{ \exp (x_i) }{ \sum_{j=1}^K \exp (x_j) },
\end{equation}
which gives the outputs of probabilistic forecasts. We determine the parameters in Table 1.

Figure 1(a) shows the architecture of DeFN \citep[see][for simplified representations]{nis18}, while Figures 1(b) and 1(c) show the architectures of DeFN-R, which were retuned to 
increase BSS for $\geq$M-class and $\geq$C-class flare predictions (see Appendix A.1). DeFN-R includes simple skip connections \citep[see also the residual network in][]{he15} 
and batch normalization represented by the notation BN \citep{iof15} to stabilize the training and improve the precision of the model. The number of nodes was investigated in the 
range of 79--200, and the batch size was in the range of 50--200. The architecture of DeFN-R with 5--12 layers was surveyed by attaching or detaching skip connections and 
dropouts. Finally, we selected the network architecture that provides representative results in the pilot experiment, as shown in Table 2.

\begin{deluxetable}{llllrl}
\tabletypesize{\normalsize}
\tablecaption{Symbol notations. \label{tbl1}}
\tablewidth{0pt}
\startdata
\tableline
\colhead{x, {\bf x}} & \colhead{Arbitrary input parameters}                                                                                         \\[+0.1cm]
\colhead{y, {\bf y}} & \colhead{Arbitrary output parameters (discrete or continuous)}                                                       \\[+0.1cm]
\colhead{N} & \colhead{Number of training samples}                                                                                                 \\[+0.1cm]
\colhead{K} & \colhead{Number of classes/categories}                                                                                              \\[+0.1cm]
\colhead{${\bf y}_n^*$ = $\{$ ${\bf y}_{nk}^*|$ \textit{n}=1,....,\textit{N} $\}$ } & \colhead{Correct label of nth training sample}    \\[+0.1cm]
\colhead{$p$(${\bf y}_n$) = $\{$ $p$($y_{nk}$)$|$ \textit{n}=1,....,\textit{N} $\}$ } & \colhead{Estimated probability of ${\bf y}^*$} \\[+0.1cm]
\enddata
\end{deluxetable}

\begin{figure}[hbtp]
\includegraphics[scale=0.9]{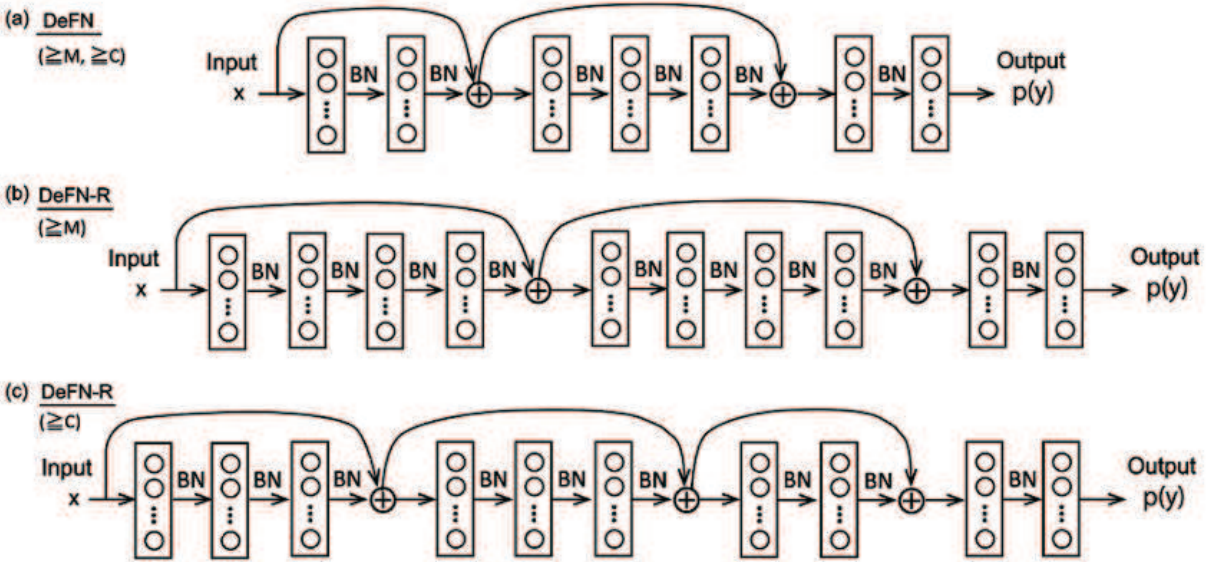}
\caption{Architectures of (a) DeFN and (b), (c) DeFN-R for $\geq$M-class and $\geq$C-class flare predictions.\label{fig1}}
\end{figure}

\clearpage
%
%
\subsection{Optimization Method for DeFN-R}

For classification problems, models are usually trained by optimizing parameters to minimize the cross entropy. The cross entropy between $p(y_k^*)$ and $p(y_k)$ is determined 
by the following equation:
\begin{equation}
J_{CE} = - \sum_{k=1}^K p(y_k^*) \log p(y_k),
\end{equation}
where we omitted $n$ representing the $n$th sample for simplicity. $p(y_k^*)$ is the initial probability of correct labels $y_k^*$, i.e., 1 or 0, while $p(y_k)$ is forecast probability. 
The components of $y_k^*$ are 1 or 0, and thus, $p(y_k^*)$ = $y_k^*$.

In DeFN, since the flare occurrence rate is imbalanced, the following summation of the weighted cross entropy was used as the loss function instead:
\begin{equation}
J_{WCE} = - \sum_{n=1}^N \sum_{k=1}^K w_k y_{nk}^* \log p(y_{nk})
\end{equation}
Here, $w_k$ is the weight of each class, which was set to the inverse ratio of the class occurrence in DeFN, e.g., [1, 50] for $\geq$M-class flare events and [1, 12] for 
$\geq$C-class flare events. However, this results in reduced reliability. Thus, in this paper, we set the weight to be constant, i.e., $w_k$ = [1, 1].

The parameters used here are summarized in Tables 1 and 2. As the method for stochastic optimization, we used adaptive moment estimation \citep[Adam;][]{kin14}, which is 
an extension of AdaGrad, RMSprop, and AdaDelta. We used the recommended values for Adam's hyperparameters, which are given in Table 2, because overfit hyperparameters 
only work with certain architectures.

\begin{deluxetable}{llllrl}
\tabletypesize{\normalsize}
\tablecaption{Parameter settings of DeFN-R ($\geq$M, C) using cross entropy as a loss function.  \label{tbl2}}
\tablewidth{0pt}
\startdata
\tableline 
\\[-0.4cm]
\colhead{(a) $\geq$M-class flares} & \colhead{}                                                                                  \\[+0.1cm]
\colhead{Loss function}           & \colhead{Summation of cross entropy}                                            \\[+0.1cm]
\colhead{Optimization method} & \colhead{Adam (learning rate = 0.001, $\beta_1$ = 0.9, $\beta_2$ = 0.999)} \\[+0.1cm]
\colhead{No. of nodes =}       & \colhead{79, 150, 150, 150, 79, 150, 150, 150, 79, 150, 2}                           \\[+0.1cm]
\colhead{Batch size =}            & \colhead{150}                                                                                   \\[+0.1cm]
\colhead{BSS =}                    & \colhead{0.298}                                                                                 \\[+0.1cm]
\tableline 
\\[-0.4cm]
\colhead{(b) $\geq$C-class flares} & \colhead{}                                                                                  \\[+0.1cm]
\colhead{Loss  function}          & \colhead{Summation of cross entropy}                                            \\[+0.1cm]
\colhead{Optimization method} & \colhead{Adam (learning rate = 0.001, $\beta_1$ = 0.9, $\beta_2$ = 0.999)} \\[+0.1cm]
\colhead{No. of nodes =}       & \colhead{79 (input), 150, 150, 79, 150, 150, 79, 150, 79, 150, 2 (output)}        \\[+0.1cm]
\colhead{Batch size =}            & \colhead{150}                                                                                    \\[+0.1cm]
\colhead{BSS =}                    & \colhead{0.412}                                                                                  \\[+0.1cm]
\enddata
\end{deluxetable}

\clearpage
%
%
\section{Results of Probabilistic Forecasts}
\subsection{Reliability Diagram}

Figures 2(a) and 3(a) respectively show reliability diagrams of probabilistic forecast results for $\geq$M-class flare events obtained by DeFN and DeFN-R. The blue line graph depicts 
the conditional expectation values of the outcome. The diagonal dotted line indicates perfect reliability, on which 95\% consistency bars are attached \citep{bro07, jol12}. The 95\% 
consistency bars show the ranges within which 95\% of the conditional expectation values of the outcome given the probability would fall if it were assumed that the original data were 
sampled from a perfectly reliable probabilistic forecast system, namely, the null hypothesis that the probabilistic forecast is perfectly reliable is rejected with the 95\% significant level 
if the sampled data located outside the consistency bars. The horizontal dotted line is the climatological event rate, and the line between the perfect reliability and the climatological 
event rate corresponds to BSS = 0. Red histograms show the number of probabilistic forecasts within bins.

Plots located above the diagonal dotted line indicate under-forecasts, while plots located below the line indicate over-forecasts. In Figure 2(a), plots are located below BSS=0 line. 
This means that DeFN lacks reliability. On the other hand, in Figure 3(a), the reliability diagram is markedly improved, achieving BSS = 0.298. Reliability diagrams for $\geq$C-class 
flare events are shown in Figures 4(a) and 5(a), and a similar improvement is shown, although the reliability for $\geq$C-class flare prediction is higher (BSS = 0.412). Larger BSS 
does not always correspond to a better reliability diagram. When BSS is improved, a reliability diagram sometimes gets worse. Since the number of forecasts is larger for a smaller 
forecast probability, it is more efficient to improve a small forecast probability; thus, the model learns to fit a lower forecast probability preferentially, and large BSS does not always 
coincide with a better reliability plot.

\begin{figure}[hbtp]
\includegraphics[scale=0.75]{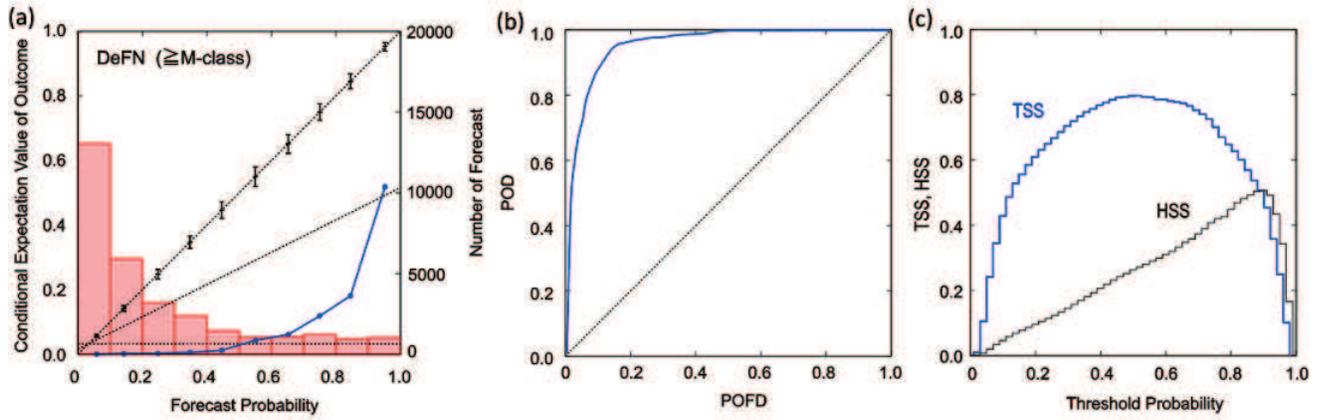}
\vspace*{+1.cm}
\caption{(a) Reliability diagram, (b) ROC curve, and (c) variability of TSS and HSS for probabilistic forecast of $\geq$M-class flares by DeFN. In panel (a), the diagonal dotted line 
indicates perfect reliability, on which 95\% consistency bars are attached. The horizontal dotted line is the climatological event rate, and the line between the perfect reliability and 
the climatological event rate corresponds to BSS = 0. Red bars show the number of probabilistic forecasts within bins. \label{fig2}}
\end{figure}

\begin{figure}[hbtp]
\includegraphics[scale=0.75]{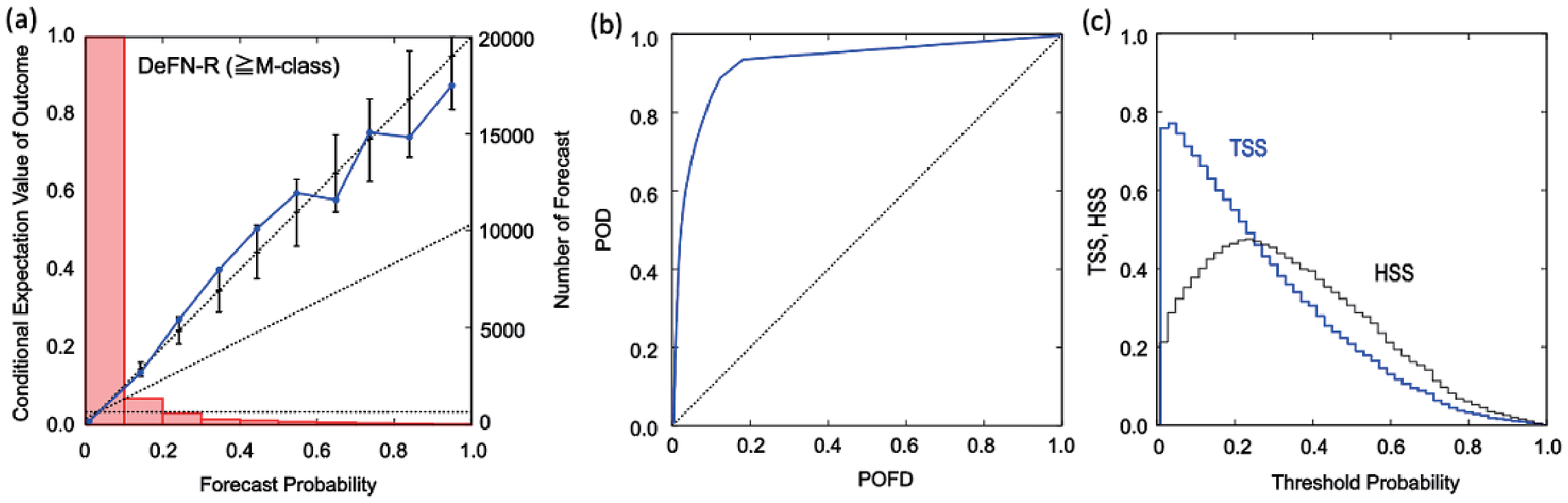}
\vspace*{+1.cm}
\caption{(a) Reliability diagram, (b) ROC curve, and (c) variability of TSS and HSS for probabilistic forecast of $\geq$M-class flares by DeFN-R. \label{fig3}}
\end{figure}

\begin{figure}[hbtp]
\includegraphics[scale=0.75]{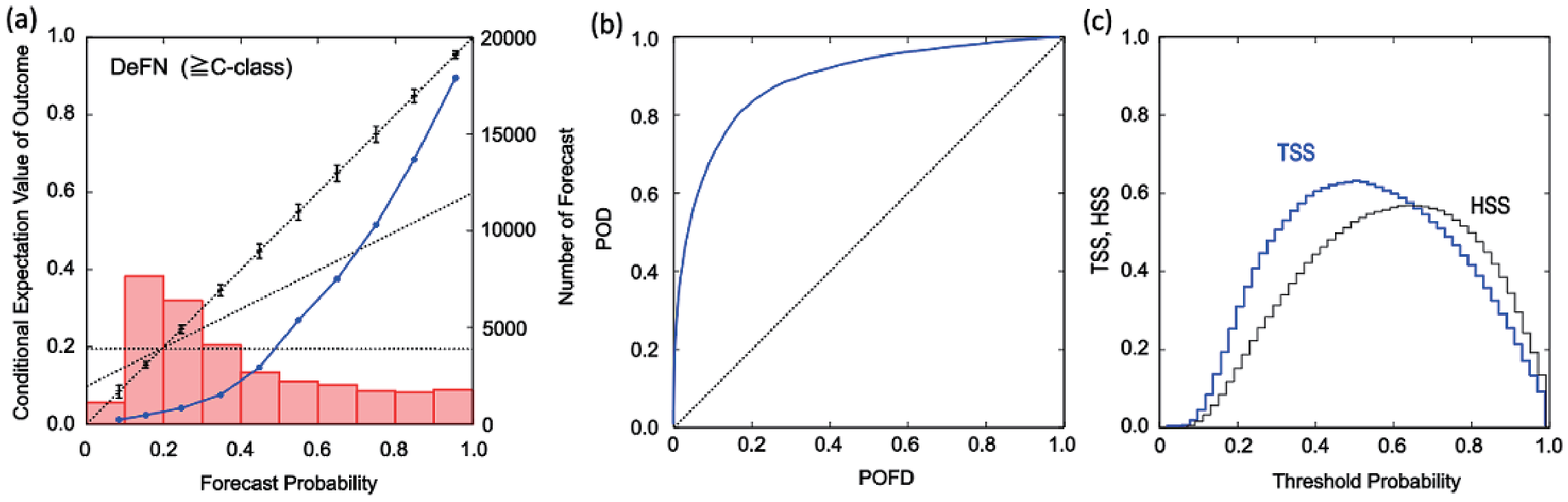}
\vspace*{+1.cm}
\caption{(a) Reliability diagram, (b) ROC curve, and (c) variability of TSS and HSS for probabilistic forecast of $\geq$C-class flares by DeFN. \label{fig4}}
\end{figure}

\begin{figure}[hbtp]
\includegraphics[scale=0.75]{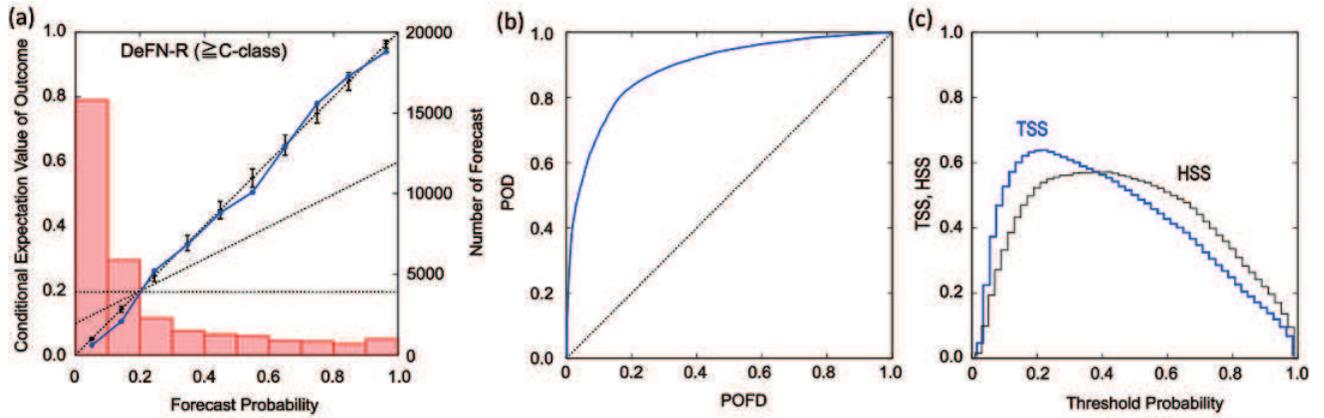}
\vspace*{+1.cm}
\caption{(a) Reliability diagram, (b) ROC curve, and (c) variability of TSS and HSS for probabilistic forecast of $\geq$C-class flares by DeFN-R. \label{fig5}}
\end{figure}

\clearpage
%
%
\subsection{Probability Threshold and ROC Curves}

\newcommand{\argmax}{\mathop{\rm argmax}\limits}

The output of the probabilistic forecast for a dataset is denoted as $p(y)$. The skill scores for deterministic forecasting require a probability threshold ($P_{th}$), above which an 
event is predicted to occur. When we set the probability threshold, the output is determined depending on whether $p(y)$ is greater than $P_{th}$ or not. In DeFN or machine 
learning algorithms trained for deterministic forecasting, the threshold is set to 50\% by default, which effectively optimizes TSS, because the errors of forecast probabilities in 
both categories are treated equally. In probabilistic forecast models such as DeFN-R, the threshold can be freely selected. In these models, categorical skill scores are calculated 
from contingency tables.

Figures 2(b) and 3(b) show the ROC curve, which tracks the performance of TSS as a function of $P_{th}$, because the two axes, probability of detection (POD) and probability 
of false detection (POFD), are the first and second terms of TSS, respectively (= POD -- POFD) (see Appendix A.2). The ROC curve shows the discrimination performance of 
correctly recognizing the flare occurrence. Note that the data with the maximum TSS are mostly away from the diagnostic line, but not the data closest to the point (0, 1). 

Figures 2(b) and 3(b) are almost the same. For these figures, the ROC score, defined as the area under the ROC curve (AUC) \citep{mas99}, is 0.96 and 0.93, and the ROC skill 
score (ROCSS), defined as ROCSS = 2AUC-1, is 0.91 and 0.86, respectively. Therefore, DeFN-R improved the reliability diagram while maintaining the ROC curve. Namely, DeFN-R 
achieved both high discrimination performance and high reliability. Similar results are also shown in Figures 4(b) and 5(b) for $\geq$C-class flare probabilistic forecasts, where 
AUC is 0.89 and 0.89, and ROCSS is 0.77 and 0.78, respectively. In other words, the ROC curve and the variability of TSS in section 4.3 are not significantly affected by small 
changes of a loss function and an architecture, while a reliability diagram and BSS considerably change.

\clearpage
%
%
\subsection{Variabilities of TSS and HSS}

TSS is a function of threshold probability ($P_{th}$), which is shown in Figures 2(c) and 3(c) for $\geq$M-class flare predictions. We took a threshold step of 0.02, resulting in 49 
data, and a grid covering the region [0, 1] excluding the edges. In Figure 2(c), TSS continuously changes with the threshold probability and peaks at around 0.50 (50\%) for DeFN. 
On the other hand, in Figure 3(c), TSS peaks at around 0.032 (3.2\%), which is the climatological event rate of $\geq$M-class flare events in the datasets. In more detail, the 
maximum values of TSS are 0.80 and 0.77 at $P_{th}$ = 0.52 and 0.04 in Figures 2(c) and 3(c), respectively. This is confirmed by similar ROC curves of DeFN and DeFN-R. Similar 
findings are obtained for $\geq$C-class flare predictions, as shown in Figures 4(c) and 5(c). In Figure 4(c), TSS peaks at around 0.50 (50\%), while it peaks near the climatological 
event rate of 0.196 (19.6\%) in Figure 5(c). The maximum values of TSS are 0.63 and 0.64 at $P_{th}$ = 0.50 and 0.22 in Figures 4(c) and 5(c), respectively.

As a result, DeFN-R for $\geq$M-class and $\geq$C-class flare predictions shows that the maximum TSS occurs when $P_{th}$ is approximately the climatological event rate. 
This is consistent with previous works by Bloomfield et al. (2012), Barnes et al. (2016), and Leka et al. (2018). It has been mathematically verified that a reliable probabilistic 
forecasting model has the peak of TSS at the probability threshold of the climatological event rate \citep{kub19}. In our case, DeFN-R is reliable, so the maximum TSS occurred 
at the climatological event rate. However, since the maximum TSS did not occur at this rate, DeFN lacks reliability. This is related to the fact that, in DeFN, the threshold is set 
to 50\% to effectively optimize TSS and that the prior distribution is normalized by the weight of the cross entropy, $w_k$.

We also show the variability of the Heidke skill score (HSS) as a function of threshold probability in Figures 2(c), 3(c), 4(c), and 5(c) for comparison. The maximum values of HSS 
are 0.51 and 0.48 at $P_{th}$ = 0.90 and 0.24 for $\geq$M-class flare prediction in Figures 2(c) and 3(c), respectively. On the other hand, it is 0.57 in both Figures 4(c) and 5(c) 
at $P_{th}$ = 0.64 and 0.42, respectively, for $\geq$C-class flare prediction. The variability of HSS is also discussed in Bobra \& Couvidat (2015) and Florios et al. (2018).

\clearpage
%
%
\section{Summary and Discussion}

We investigated the reliability performance of DeFN by changing the optimization methods of machine learning and developed a reliable probabilistic forecast model of solar flares, 
named DeFN-Reliable (DeFN-R). This model is trained to predict the occurrence probabilities of flares occurring within the following 24 h after observing an image in an operational 
setting. We compared two models with different loss functions, i.e., different weighted cross entropies with $w_k$ = [1, 50] and [1,1] for $\geq$M-class and $<$M-class/non-flare 
events and with $w_k$ = [1, 12] and [1,1] for $\geq$M-class and $<$M-class/non-flare events. Then, the prediction results were evaluated using reliability diagrams, ROC curves, 
and TSS variability as a function of threshold probability.

Comparing the two models for both $\geq$M-class flare events and $\geq$C-class flare events, we found that the reliability performance of DeFN-R varies considerably. As in 
DeFN, adopting weighting in the prior distribution is reasonable for a deterministic forecast, because the errors of forecast probabilities in both categories are treated equally with 
a threshold of 50\%. On the other hand, when we compared the two models with and without weighted loss functions, the ROC curve did not change significantly, i.e., the maximum 
TSSs were almost the same. This indicates that, although adopting weighting on a loss function is effective for increasing TSS, the machine learning model itself is superior for 
prediction.

DeFN-R markedly improved reliability while keeping the ROC curve almost the same. As mathematically predicted, for reliable DeFN-R, the maximum TSS occurs at the climatological 
event rate. Although a DNN model using cross entropy as a loss function was considered, the reliability of the model has not been discussed. Regardless of whether or not weighting 
is adopted for the cross entropy, the two models are mathematically similar, although the accuracy of prediction may depend on the purpose. When we set the probability threshold 
at 3.6\% (19.6\%) for $\geq$M-class ($\geq$C-class) flare prediction, DeFN-R shows almost the same performance as DeFN. Finally, a merit of DeFN-R is that the probability 
threshold can be selected according to users' purposes, and DeFN-R has larger degree of freedom than deterministic forecasting models.

\clearpage
%
%
\appendix
\section{Appendix: Evaluation Method}
\subsection{Definition of Brier Skill Score for Probability Forecasting}

In this paper, we used the Brier skill score (BSS) for evaluation, which is a measure of probability forecasting models. BSS is calculated from the Brier score (BS) and climatological 
Brier score (BS$_c$) as follows:
\begin{eqnarray}
BSS  & = & \frac{BS - BS_c}{0 - BS_c},   \\
BS    & = & \sum_{n=1}^N (p(y_{n}) - y_{n}^*)^2,   \\ 
BS_c & = & \sum_{n=1}^N (f_i - y_{n}^*)^2           (\textit{i} = 1, 2), 
\end{eqnarray}
where $p$($y_n$) is the predicted probability and $f_i$ is the climatological event rate. Here category $k$ is fixed to 1, and $y_n$ is a value for N samples. In this paper, we evaluate 
our model for two problems: $\geq$M-class and $<$M-class, or $\geq$C-class and $<$C-class. For $\geq$M-class and $\geq$C-class flare events, $f_1$=0.032 and $f_2$=0.196, 
respectively. Note that probabilistic forecasting never achieves BS=0; thus, even perfect reliability cannot achieve BSS=1. BSS=1 can be achieved only by deterministic forecasts.

\clearpage
%
%
\subsection{Definitions of Skill Scores for Deterministic Forecasting}

We used some skill scores to evaluate our deterministic forecasting of flares, by setting a probability threshold. The probability of detection (POD) or the recall, the probability of 
false detection (POFD), the true skill statistic (TSS), also known as the Peirce skill score or the Hanssen and Kuipers discriminant, and the Heidke skill score (HSS) were used and 
determined in the following way \citep{han65, mur93, bar09, kub17}:
\begin{eqnarray}
POD  & = & \frac{TP}{TP + FN}, \\
POFD & = & \frac{FP}{FP + TN}, \\ 
TSS & = & \frac{TP}{TP + FN} - \frac{FP}{FP + TN}, \\
S & = & \frac{a+c}{a+b+c+d}, \\
HSS = \frac{PC - PC_r}{1 - PC_r} & = & \frac{2S(1-S)(POD-POFD)}{S+S(1-2S)POD+(1-S)(1-2S)POFD}, \\
PC   = \frac{a+d}{a+b+c+d} & = & S \cdot POD+(1-S) \cdot POFD, \\
PC_r & = & \frac{(a+c)(a+b)+(b+d)(c+d)}{(a+b+c+d)^2}.
\end{eqnarray}
Here, $a, b, c, d$ indicate components of a contingency table for dichotomous forecast: hits, false alarms, misses, and correct rejections, respectively.

\clearpage
%
%
\subsection{Robustness of DeFN-R}

In this paper, we utilized a single chronological split of a database for operational evaluation, namely, datasets in 2010 - 2014 for training and those in 2015 for testing (case 0). However, 
it is not enough to assess the stochastic nature of DeFN-R. Although the time-series cross-validation approach can generally give optimistic assessments of a forecasting method's 
performance, it allows multiple runs to be made, so we computed standard deviations of all the skill scores as a function of the probability threshold.

Here, we considered a few more splits, in the chosen operational settings approach. We considered the following cases: (1) 2010 - 2013 for training and 2014 - 2015 for testing, (2) 
Jun 2010 - Mar 2014 for training and Apr 2014 - Dec 2015 for testing, (3) Jun 2010 - Jun 2014 for training and Jul 2014 - Dec 2015 for testing, (4) Jun 2010 - Sep 2014 for training 
and Oct 2014 - Dec 2015 for testing, (5) Jun 2010 - Mar 2015 for training and Apr 2015 - Dec 2015 for testing, and (6) Jun 2010 - Jun 2015 for training and Jul 2015 - Dec 2015 for 
testing. Although the time-series cross-validation is more reasonable for model selection, this is out of scope.

The prediction results are shown in Table 3 and Figures 6 and 7. When the size of the training dataset was reduced in cases 1 - 3, BSS decreased as shown in Table 3. This is probably 
because the solar maximum occurred in 2014, and it appears that we had insufficient training samples when we reduced the number of samples from the datasets in 2014. In cases 4 and 
5, BSS was almost the same as that of case 0 except for case 5 for $\geq$M-class events. In case 6, BSS increased.

Figures 6 and 7 show the results for cases 0 - 6 described above in the form of reliability diagrams, ROC curves, and TSS vs probability threshold curves and for $\geq$M or $<$M-class 
and $\geq$C or $<$C-class flare predictions of DeFN-R. Cases 1, 2, and 5 for $\geq$M-class flare prediction show a lower reliability than the other cases, as can be seen in Figure 6(a). 
The reliability diagrams in Figures 6(a) and 7(a) vary with the case, mainly in the range of 0.4 - 1.0, but the variation is almost within the 95\% consistency bars in Figures 3(a) and 5(a). 
Therefore, our prediction results in case 0 obtained using our database are reasonably robust.

\begin{deluxetable}{llllrl}
\tabletypesize{\normalsize}
\tablecaption{BSS in cases 0 - 6 for $\geq$M-class and $\geq$C-class flare predictions by DeFN-R. \label{tbl3}}
\tablewidth{0pt}
\startdata
\tableline 
\\[-0.4cm]
\colhead{} & \colhead{$\geq$M-class flares} & \colhead{}  & \colhead{$\geq$C-class flares} & \colhead{}  \\[+0.1cm]
\colhead{} & \colhead{BSS} & \colhead{climatology}  & \colhead{BSS} & \colhead{climatology}  \\[+0.1 cm]
\colhead{Case1 (test: 2014 Jan -)}    & \colhead{0.2574}   & \colhead{0.0353} & \colhead{0.3530}  & \colhead{0.2161}  \\[+0.1cm]
\colhead{Case2 (test: 2014 Apr -)}    & \colhead{0.2337}   & \colhead{0.0318} & \colhead{0.3609}  & \colhead{0.2124}  \\[+0.1cm]
\colhead{Case3 (test: 2014 Jul -)}    & \colhead{0.2529}   & \colhead{0.0327} & \colhead{0.3738}  & \colhead{0.2038}  \\[+0.1cm]
\colhead{Case4 (test: 2014 Oct -)}  & \colhead{0.2849}   & \colhead{0.0350} & \colhead{0.4110}  & \colhead{0.1982}  \\[+0.1cm]
\colhead{Case0 (test: 2015 Jan -)}    & \colhead{0.2977}   & \colhead{0.0324} & \colhead{0.4107}  & \colhead{0.1958}  \\[+0.1cm]
\colhead{Case5 (test: 2015 Apr -)}    & \colhead{0.2559}   & \colhead{0.0309} & \colhead{0.4268}  & \colhead{0.1954}  \\[+0.1cm]
\colhead{Case6 (test: 2015 Jul -)}    & \colhead{0.3017}   & \colhead{0.0350} & \colhead{0.4558}  & \colhead{0.1804}  \\[+0.1cm]
\enddata
\end{deluxetable}

%

\begin{figure}[hbtp]
\includegraphics[scale=0.75]{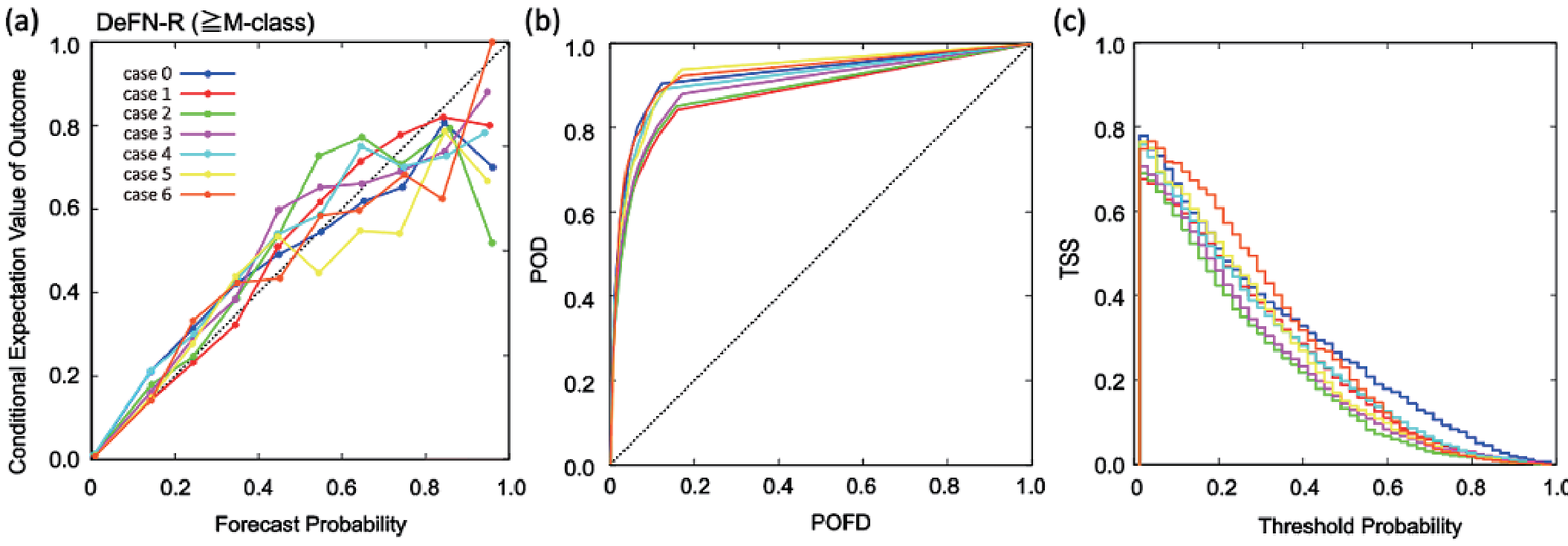}
\vspace*{+1.cm}
\caption{(a) Reliability diagrams, (b) ROC curves, and (c) variability of TSS for probabilistic forecast of $\geq$M-class flares by DeFN-R in 
cases 0 - 6. \label{fig6}}
\end{figure}

\begin{figure}[hbtp]
\includegraphics[scale=0.75]{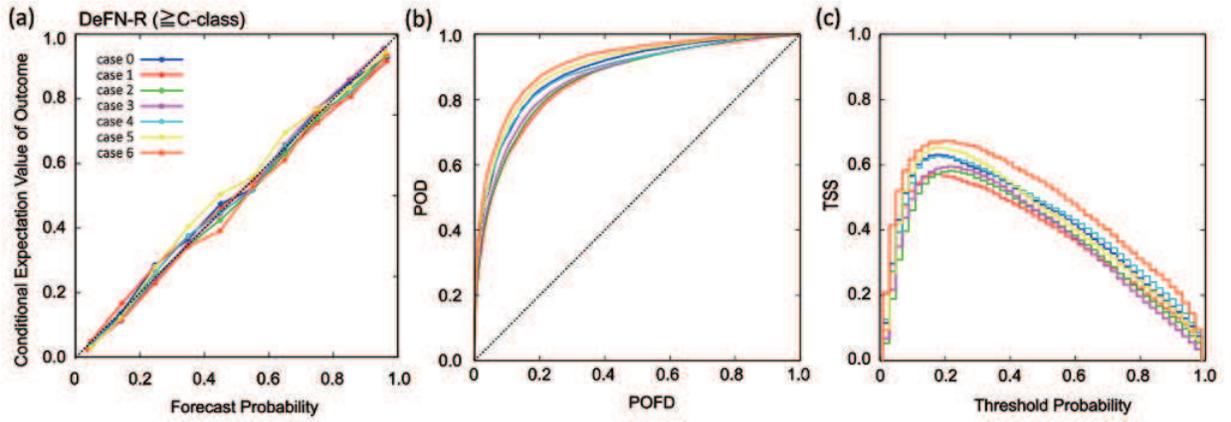}
\vspace*{+1.cm}
\caption{(a) Reliability diagrams, (b) ROC curves, and (c) variability of TSS for probabilistic forecast of $\geq$C-class flares by DeFN-R in 
cases 0 - 6. \label{fig7}}
\end{figure}

\clearpage
\acknowledgments
The data used here are courtesy of NASA/SDO, the HMI \& AIA science teams, Joint Science Operations Center (JSOC), as well as GOES team. 
This work was supported by KAKENHI grant Number JP18H04451.\\

%
%


\clearpage

%
%
%
\end{document}